\documentclass[12pt,preprint]{aastex}

\newcommand{\etal}{et~al.}

\newcommand{\msun}{M$_{\sun}$}

\newcommand{\vsini}{$v \sin i$}

\received{2006 February 20}
\begin{document}

\title{Stellar Rotation: A Clue to the Origin of High Mass Stars?}

\author{S.\ C.\ Wolff, S.\ E.\ Strom, D.\ Dror}
\affil{National Optical Astronomy Observatory, 950 N. Cherry Ave., Tucson, AZ, 85719 (swolff@noao.edu)}

\author{L.\ Lanz}
\affil{University of Maryland, College Park, MD 20742}

\and
\author{K.\ Venn}
\affil{University of Victoria, Victoria BC V8W 2Y2}

\begin{abstract}

We present the results of a study aimed at assessing whether low and high mass stars form similarly. Our approach is (1) to examine the observed projected rotational velocities among a large sample of newly-formed stars spanning a range in mass between 0.2 and 50 {{\msun}}; and (2) to search for evidence of a discontinuity in rotational properties that might indicate a difference in the stellar formation process at some characteristic mass. Our database includes both recently published values of {\vsini} for young intermediate- and low- mass stars in Orion, as well as new observations of O stars located in young clusters and OB associations. We find that the median of the quantity $v_{obs}/v_c$ (observed rotational speed/equatorial breakup velocity) is typically about 0.15 and shows no evidence of a discontinuity over the full range of stellar masses, while the quantity {Jsini/M} (derived angular momentum per unit mass) exhibits a slow, monotonic rise (J/M $\sim$ $M^{0.3}$) with increasing mass with no evidence of a discontinuity. We suggest that these observations are most simply interpreted as indicative of a single stellar formation and angular momentum regulation mechanism: one that results in rotation rates well below breakup, and angular momenta per unit mass that differ systematically by no more than a factor of 3-4 over a mass range spanning a factor of 250.

\end{abstract}


\section{Introduction}

There are two classes of models proposed for the formation of massive stars (e.g., Krumholz 2005). The first posits that massive stars are assembled in a manner analogous to their lower mass counterparts: from infalling material located in a rotating protostellar core channeled starward through a circumstellar accretion disk, with the flow of material from the inner regions of the accretion disk toward the stellar surface mediated by a magnetic field rooted in the star. We choose the shorthand ``magnetospherically-mediated accretion" (henceforth MMA) to describe this scenario.  Bonnell, Vine, \& Bate (2004) propose the following variant: that massive stars form preferentially in cluster-forming environments, starting their lives as lower mass cores whose mass then increases as a result of competitive accretion of surrounding molecular gas. In the case of mass buildup through competitive accretion, transfer of material from the protostellar core to the central forming star would presumably still follow the general precepts of MMA.  A second, very different class of model posits that massive stars start their lives as lower mass protostellar cores in clusters, and then grow via mergers of cores located in regions where the density of cores is high  (Bally \& Zinnecker 2005 and references therein). We refer to this class of model as  ``core merger " models or CM. Observational tests aimed at determining which of these contrasting models, MMA or CM, dominates in nature have proven impossible to date, primarily because high mass stars form on very short time scales in highly obscured regions at distances large enough so that, with the spatial resolutions achieved by current generation telescopes, crowding renders study of individual forming objects difficult.   

Direct tests of MMA for massive stars would involve measurements similar to those carried out to establish MMA as the likely assembly mechanism for low mass stars: of stellar magnetic fields (e.g. Johns-Krull, Valenti, \& Koresko 1999), and of line profiles diagnostic of the temperature-density-velocity fields expected were material channeled starward from the inner regions of the accretion disk by magnetic fields along ``funnel flows" (e.g., Hartmann, Hewett, \& Calvet, 1994; Muzerolle, Calvet, \& Hartmann 1998). Unfortunately, rotational and turbulent broadening among higher mass stars precludes direct observation of Zeeman broadening for magnetically sensitive lines (rotational broadening  typically exceeds Zeeman broadening  by more than a factor of 10). Moreover, searches for line profile morphologies consistent with magnetic funnel flows have yielded no convincing evidence of such flows for stars with masses much in excess of 3 {\msun} (e.g., Muzerolle \etal\ 2004).  An alternative, but more indirect test of MMA might be provided by observation of highly collimated outflows presumably launched in the inner regions of the accretion disk, perhaps near the point where the dynamical pressure of accretion is matched by the magnetic pressure of the stellar field (e.g., Shu \etal\ 1995).  Such collimated outflows are common among low mass stars ($M < $ 3 {\msun}) which exhibit funnel flow profiles and for which direct measurements of magnetic field strengths are available, that is, objects where MMA finds strong observational support. Collimated outflows have also been observed in association with even more massive (up to $\sim$10 {\msun}) stars surrounded by circumstellar accretion disks (Cesaroni \etal\ 2005; Beuther \& Shepherd 2005, and references therein), perhaps suggesting that MMA is operative for stars as massive as 10 {\msun}.

Bally \& Zinnecker (2005) propose two possible tests for CM. The first involves surveys aimed at detecting direct evidence of collisions and mergers of low mass protostellar cores via observation of infrared ``flares". The second is indirect, namely a change in the character of stellar outflows from the highly collimated outflows characteristic of lower mass objects  ($M <$ 10 {\msun}) presumably undergoing MMA, to much more poorly collimated outflows among higher mass objects. Unfortunately, no surveys aimed at detecting flares in regions of high mass star formation have yet been carried out. Moreover, there is thus far only very limited evidence regarding the degree of collimation, or lack thereof, for outflows from forming stars with masses larger than $\sim$ 10 {\msun}.  

Given the challenge of distinguishing observationally between MMA and CM models for high mass star formation, alternative clues are potentially valuable.  One such clue  may be provided by observation of stellar  angular momenta among young, high mass stars.  For example, Bally \& Zinnecker (2005) suggest that the product of the merger of two stars should be rotating rapidly: a result of converting  the orbital angular momenta of merging  protostars into the spin angular momentum of  the resulting higher mass, merged object.  In contrast, stars formed via MMA should be rotating relatively slowly, at speeds well below breakup (e.g., Shu \etal\ 1994).

In a recent paper discussing the rotation properties among young stars in Orion, Wolff, Strom, \& Hillenbrand (2004; hereafter WSH) used observations of {\vsini}\ to derive values of the projected specific angular momentum ({Jsini/M}) for pre-main-sequence stars still on convective tracks; these stars have recently completed their main accretion phase and should provide the best directly measurable estimate of the initial values of angular momentum.  WSH found that the upper envelope of {Jsini/M} varies slowly with mass (J/M $\sim$ $M^{0.25}$) over the mass range 0.1-3 {\msun} and lies well below the angular momentum associated with stars having rotation speeds near breakup (see section 3 below).  These authors suggest that MMA can in principle account for these results for stars in this mass range.

In this paper we extend measurements of {\vsini} to O and B0 stars in extremely young clusters and associations and ask whether there is a discontinuity in rotation properties between lower mass stars and stars with masses $M >>$ 3 {\msun}, as might be expected if  MMA dominates the formation of low mass stars, while
CM plays a role in the formation of a significant fraction of high mass stars.

\section{Observations}
\label{sec:obs}

The O stars for this study were selected from the associations studied by
Massey \& Thompson (1991), Hillenbrand \etal\ (1993), and Massey, Johnson, 
\& DeGioia-Eastwood (1995).  Because our proposed search for a discontinuity 
in rotation properties as a function of stellar mass requires our estimating 
rotation rates for stars as soon as possible after they have formed, we have 
limited our sample to stars very close to the zero-age main sequence (ZAMS), 
i.e., to luminosity classes III-V. The stars were observed with the Hydra 
multi-object spectrograph at the 
WIYN telescope on Kitt Peak.  

Observations (the ``high resolution sample'') of nine O stars in NGC 6611 were 
obtained in May, 2005 using the Hydra multiobject fiber spectrograph on WIYN as part of a more extensive study of stellar rotation in this cluster. In conjunction with an Echelle grating and an order separating filter, these observations with WIYN-Hydra yielded spectra with R $\sim$ 20,000 spanning the wavelength range 4450 to 4590 $\AA$. In addition to our NGC 6611 program stars, 12 stars with spectral types from O4 to B0 with known values of {\vsini} (Penny 1996; Howarth et al. 1997) were observed on the same nights with the same instrumental configuration to serve as rotation standard stars. Our approach was to measure the FWHM of He I 4471 $\AA$ and to derive a relationship between FWHM and published {\vsini} values (Penny 1996; Howarth \etal\ 1997) for the twelve standard stars.  We note that the {\vsini} values adopted from these two published studies result from application of a cross-correlation technique to the rich near-ultraviolet metallic spectra of late-O and early B- stars observed with IUE.  By using stars 
from these two previous studies as standards, we
are placing our own values of {\vsini} on a known system.  We note that 
different calibrations yield slightly different answers.  For examples, 
the Penny and Howarth et al. values are larger by about 10\%  than the 
values determined from visual examination of observed line profiles by Conti \& Ebbets (1977) from spectra in the optical region of the spectrum. In turn, these latter values are systematically smaller by about 30 km/sec than the values 
derived in the pioneering study of Slettebak (1956).  

Forty-two additional stars were observed over 
the wavelength range 4070-4580 $\AA$ at a resolution of $\sim$ 0.75 $\AA$ 
(``low resolution sample''). These observations enable measurement of 
projected rotational velocities, {\vsini} $>$ 50 km/sec, and upper limits to {\vsini} for stars rotating more slowly than 50 km/s.  The spectra were 
extracted and calibrated in wavelength by Kim Venn. The lines used to
measure {\vsini} were He I 4471 $\AA$ or He I 4387 $\AA$ or both, depending
on the spectral type and the strength of the lines.  

In order to calibrate {\vsini} for our ``low resolution sample'', we established the relationship between FWHM and {\vsini} for He I 4471 $\AA$ and He I 4387 
$\AA$ by using the {\vsini} values determined as above from the 9 O stars
in NGC 6611 that were observed at both resolutions. To strengthen the 
calibration, we also added 4 stars in 
NGC 6611 with spectral types B0.5 that were also observed at both high and 
low spectral resolutions; the values of {\vsini} for these 13 stars range 
from 30 to 400 km/sec.  (The measurements of stars in NGC 6611 with types B0.5 and later will be reported along with data for B stars in several other associations in a subsequent paper.)  Independent measures of the line widths for stars observed more than once during our run indicate that the values of {\vsini} are internally consistent to within about 20 km/sec.  Given the uncertainties in the calibration and the difficulty of measuring extremely broad lines, we report the overall uncertainty in a measured value of {\vsini} as $\sim$10\% or 20 km/sec, whichever is larger.  This uncertainty is typical of {\vsini} studies for stars in this spectral type range. For example, the average difference in the {\vsini} values measured by both Penny (1996) and Howarth \etal\ (1997) for the 9 of our standard stars that are common to both of the published studies is 19 km/sec.  As a further check on our values, we have compared our results with 11 stars for which data are also available on either WEBDA (http://www.univie.ac.at/webda/webda.html) or from Huang \& Gies (2006).  We find that the standard deviation of the difference between our results and the literature values is 13 km/sec.

The stars observed, along with spectral types, etc., are listed in Table 1.  The first column lists the name of the cluster or association, the second gives the name of the star, if any; the third lists the number on the WEBDA system; the fourth and fifth columns give  the stellar position; the sixth lists the spectral type, the seventh the log of the effective temperature, the eighth the bolometric luminosity, the ninth the stellar mass, the tenth the apparent rotational velocity, and the eleventh the log of the specific angular  momentum. The derivation of the quantities in the last 5 columns is described in section 3.1

\section{Analysis}

\subsection{Comparing Observed Rotational Velocities with Equatorial Breakup Velocities}

In order to make our first assessment of whether forming stars of all masses rotate similarly or whether there is a discontinuity at some characteristic mass, we need to establish a simple procedure for comparing rotation properties of young high and low mass stars.   Our first approach is to compare the observed rotational velocities of the stars in our sample with the critical velocity for stars of the same mass on the birthline.  We will argue below that the observed values of {\vsini} for the very young stars in our sample have changed little since the time that the main accretion phase ended, and so the current values of {\vsini} are representative of their initial values.  The goal of this comparison is to determine whether the ratio of the observed initial values of {\vsini} to the critical velocities on the birthline varies systematically with mass.  This simple approach is motivated by the fact that explicit calculation of stellar angular momentum, which is a more fundamental quantity, requires a number of assumptions about stellar structure and the internal distribution of angular momentum as well as accurate estimates of stellar parameters.  The apparent rotational velocity, however, is a directly measured quantity, while the breakup velocity on the birthline can be estimated from stellar models.  Therefore, we begin the analysis by comparing the observed values of {\vsini} with the critical velocities calculated from models of newly formed stars.

Models predict that stars with masses comparable to the O-stars in our sample are deposited directly on the zero age main sequence at the end of the assembly process.  For a time-averaged mass accretion rate of $dM_{acc}/dt$ = 10$^{-5}$ {\msun}/year, which observations suggest is typical for intermediate mass stars (Palla \& Stahler 1992), stars with M $>$ $8$ {\msun} should lie on the main sequence when accretion stops.  If $dM _{acc}/dt $ =10$^{-4}$ {\msun}/year, which is more typical of the accretion rates advocated by McKee \& Tan (2003) for 20-30 {\msun}  stars, then stars with $M >$ 15 {\msun} should lie on the main sequence when accretion stops.  Therefore, we expect that all of the O stars in our sample were already on the main sequence when accretion ended.
  
For condensed stasr, such as O stars on the main sequence,  the breakup velocity $v_c$ is given by (Townsend, Owocki, \& Howarth 2004)
	$v_c$ = $(2GM/3R_p)^{0.5}$,
where M is the mass of the star and $R_p$ is the polar radius at the
breakup velocity. The factor 2/3
takes into account the difference in the polar and equatorial radii for the
case of critical rotation.
For the O stars, most authors calculate the critical velocity by including
a correction factor for radiation pressure:
	$v_c^{2}$ = $(GM/R_e)(1-\Gamma_{rad})$
where $\Gamma_{rad}$, the ratio of the luminosity to the Eddington luminosity is given by Mihalas (1978): 
	$\Gamma_{rad}$ = 2.6 x $10^{-5}$ L/M.
Maeder and Meynet (2000) point out that this relationship is valid only if we
assume that the brightness of a rotating star is uniform over its surface,
which is inconsistent with von Zeipel's theorem.  They argue that radiation
pressure can in fact be ignored for stars with Eddington factors less than
0.639, which is true for the stars treated here. 
For present purposes, we will also assume that the polar radius of a rapidly rotating star can be taken to be approximately equal to the radius of a non-rotating star.  We have used the models of Schaller \etal\ (1992) to obtain the parameters for ZAMS O stars.  The critical velocities calculated with no allowance for radiation
pressure agree well with the calculations by Meynet \etal\ (2006). 

For an accretion rate of $10^{-5}$ {\msun}/yr, stars with masses less than 8 
{\msun} have not yet reached the ZAMS when
the main accretion phase ends.  To estimate the critical rotational velocities 
for these lower mass stars when accretion stops and they are deposited on the birthline, we have used the PMS models of Swenson \etal\ (1994) and the birthline of Palla and Stahler (Palla, private communication) for an accretion rate of $10^{-5}$ {\msun}/yr to determine the values of M and R on the birthline.  The Swenson models extend 
only to 5 {\msun}.  We have used the models of Siess, Dufour, \& Forestini 
(2000) to calculate values for stars of 3-7 {\msun}; there is good agreement in the overlap region.  

We have been unable to find calculations of the critical
velocity derived from models of PMS stars.  However, it appears that 
the 2/3 factor used
for high mass stars is a reasonable first approximation even for PMS stars.  
Herbst and Mundt (2005)
have used the equilibrium shapes of a rotating polytrope of index 1.5 
to estimate that the ratio of the polar to equatorial radius for the most 
rapidly
rotating ($P=0.6$ days) low-mass PMS stars is $\sim$ 0.75.  We have also derived
the true rotational velocities for the 5 PMS stars in Orion with periods 
less than 0.9 days (Stassun et al. 1999) and masses between 0.15 and 0.27
{\msun}.  If we assume that these stars are rotating at breakup, we find that
the lower bound for the factor that should be used to calculate
the critical velocity is 0.62.  Because these values bracket the factor
of 2/3 calculated for high mass stars, we have assumed that this same factor 
can also be applied to low-mass PMS stars.

The results are plotted in Figure 1.  For low and intermediate mass stars, we have used the values of {\vsini} derived by Rhode, Herbst \& Mathieu (2001) and WSH for stars in the Orion association, most of which are less than $10^6$ years old. (We note that rotation periods have been derived from photometry of many more low mass stars in Orion but cannot be measured for the O stars, which lack the spot-modulated light curves characteristic of their lower mass brethren.   Therefore, we choose here to confine our comparison to {\vsini}, which can be measured for stars of all masses.).  

For the high mass stars in our current sample, we use the values of {\vsini} reported in Table 1.  We derive masses for the O stars from the evolutionary tracks of Schaller \etal\ (1992).  In order to do so, we need to know the effective temperature ($T_{eff}$) and bolometric luminosity ($L_{bol}$).  We have used the calibrations given by Massey \etal\ (2005) to estimate $T_{eff}$ from spectral types given in the papers by Massey \& Thompson (1991), Hillenbrand \etal\ 1993, and Massey \etal\ (1995); given $T_{eff}$, the bolometric correction can also be obtained from Massey \etal\ (2005).  The distances to each association have been taken from the papers that provided the spectral types, and intrinsic colors as a function of spectral type given by Fitzgerald (1970) have been used to derive to the reddening to each star. 

Figure 1 shows that nearly all of the stars are rotating at velocities that are less than half the critical velocity on the birthline, and most are rotating at velocities that are less than 30\% of the critical velocity.  We note that in 
this figure, we have plotted $v_c$ but have measured {\vsini}.  A statistical
correction can be made for a group of stars whose inclinations are randomly
distributed by multiplying the average {\vsini} by $4/\pi$ to obtain
the average true velocity, which we will call $v_{obs}$ (Chandrasekhar \& 
Munch 1950; Gaige, 1993).   
If we subdivide the low mass stars into three groups according to mass, we find that the median value of $v_{obs}/v_c$ $\sim$ 0.14 for the lowest mass group, 
0.10 for next lowest mass group, and 0.11 for stars with masses near that of the Sun (see Table 2). If we divide the high mass stars into two groups, the medians are 0.13 for stars with masses between 8 and 25 {\msun} and 0.20 for the stars with masses greater than 25 {\msun}. 

The discerning reader will note a ``gap" in the observed data between 3 and 8 {\msun}. This gap results from the fact that, while there are many observations of stars near the ZAMS in this mass range, there are very few observations of stars young enough to be near their expected birthline radii. Rotation rates for ZAMS stars in this mass range may therefore not be representative of their initial values. Rather, such stars are initially deposited on pre-main sequence radiative tracks and will spin up as they contract, resulting in both higher critical velocities and higher rotation rates on the main sequence than would have been
observed on the birthline.  

The results from our limited sample appear to be representative of the results from other studies.  Stassun \etal\ (1999) also find only a few exceptional low mass PMS stars in Orion with periods close to the critical value, and our results for the low mass Orion sample plotted in Figure 1 are representative of the range of apparent rotational velocities observed for other groups of PMS stars of similar mass.  More extensive studies of O stars have found that about 95\% of main sequence O stars have values of {\vsini} that are less than 300 km/sec (Penny 1996; Howarth \etal\ 1997), with a tail extending to about 400 km/sec.  Because the effects of limb darkening are severe in stars rotating near the breakup velocity, model calculations for B stars show that line widths may become insensitive to rotation speeds higher than about 90\% of the critical velocity for stars viewed equator-on (Townsend \etal\ 2004).  Only about 5\% of the O stars, however, are rotating even as rapidly as half the critical velocity, and so our sample should not be affected by excessive limb darkening.

Was the rotation of our sample stars also low relative to the critical velocity immediately after accretion stopped?  Or has significant angular momentum loss occurred since the stars were deposited on the birthline or, equivalently for the high mass stars, since they first arrived on the ZAMS?  

For stars with masses $M >$ 15 {\msun}, stellar winds are predicted to carry away angular momentum following deposition on the ZAMS. Moreover, as massive stars evolve, the radius increases, thus driving surface rotation speeds toward lower values. The amount of angular momentum lost depends on the age of the star,
and unfortunately it is difficult to estimate the ages of massive stars with
an accuracy of 1-2 Myr.  The 
values of T$_{eff}$ and L$_{bol}$ derived from the non-LTE atmospheric models
displace even the youngest O stars to the right of the ZAMS predicted by 
models of the 
interior (Schaller et al. 1992) by 2000-4000 K (Repolust, 
Puls, \& Herrero 2004; Martins, Schaerer, \& Hillier 2005).  As an 
alternative method of estimating ages, 
we note that there is evidence that all of our clusters except CygOB2 contain 
PMS stars
with masses of 3-5 {\msun} and temperatures cooler than T$_{eff}< 10000$K.
Stellar models (e.g. Siess \&Forestini 2000) predict that these stars are 
less than 2.6 Myr old.  Since the formation of stars in these regions appears to
be contemporaneous within 2-3 Myr with the most massive stars being the 
youngest (cf. Massey et al. 1995), it is reasonable to assume that the stars in 
our sample are less than 2.5 Myr old.  The observations of Cyg OB2 do not go 
faint enough to reach the PMS stars, but Massey et al. have estimated from 
the massive stars that the age of Cyg OB2 is similar to that of the other 
associations in our sample.  

Only 4 of the stars in our sample have masses greater than 40 {\msun}.  
Quantitative estimates from extant models predict that the surface rotation rate will be reduced by only about 1/3 of the initial value during the first 2.5 Myr for stars rotating at 300 km/sec and with masses of ~40 {\msun} (Meynet \& Maeder 2000).  For most of the stars in our sample, which are of lower mass, the 
effects of winds and radius changes will be smaller.  Therefore, we can assume that the currently observed values of {\vsini} for the O stars in our sample differ by not more than 100 km/sec, and typically by much less, from the initial
values just after the star was fully assembled.  For stars more massive
than 40 {\msun}, the angular momentum loss is predicted to be larger and,
given the uncertainty 
in ages, we can say little about what their initial angular momentum might 
have been.

Some low mass stars must lose angular momentum as they evolve toward the ZAMS along convective tracks (Rebull, Wolff \& Strom, 2004; Herbst \& Mundt, 2005).  For this reason, the values of {\vsini} for our sample stars could in principle be somewhat lower than their initial values when they were deposited on the birth line.  For example, Covey \etal\ (2005) report that Class I/flat spectrum objects rotate on average $\sim$ 38 km/sec, or about twice as fast on average as Classical T Tauri stars.  An increase in the average rotation rate of a factor of two would, however, still result in a representative median value of $v_{obs}/v_c \sim 0.2$. 

What we can conclude from the data in Figure 1, therefore, is that throughout the mass range from 0.1 to 50 {\msun}\  the median value of $v_{obs}$ is 
$\sim$ 10-20\%  of the critical velocity; this difference does not change significantly with mass.

\subsection{Comparing Stellar Angular Momenta for Stars of Differing Mass}

A more fundamental quantity than rotation speed is the angular momentum per unit mass.  
The specific angular momentum of a star is given by
	J/M = I$\Omega$/M.
Therefore, in order to calculate the specific angular momentum of a star we require values for the moment of inertia (I), the angular rotational velocity, the stellar radius, and the mass.  We also assume that the observed surface rotation rate is representative of the rotation of the star as a whole.

Most published models do not provide moments of inertia, but Maeder (private communication) has provided us with values for models of stars with masses of 3, 5, 9, and 15 {\msun} on the ZAMS. Meynet \etal\ (2006) have shown that log I scales as log M for massive main sequence stars, and we have used this relationship to extrapolate the values of I from the models we have to the higher mass stars.

For the highest mass star for which we have a model (15 {\msun}), the computed tracks predict a change of log I of only 0.11 dex  by the time that such a star has completed nearly half its evolution toward  core hydrogen exhaustion.  Since this is a small change, we neglect evolutionary effects in I for the relatively unevolved stars in our sample.

The calculated values of J/M are plotted as a function of mass in Figure 2. As seen in this Figure, J/M varies slowly with mass over the range 0.2-50 {\msun}.  A fit by eye suggests that over this entire range the upper bound to  $J/M \sim M^{0.3}$, which is very close to the value of 0.25 derived by WSH for stars with $M <$ 3 {\msun}.  The scatter below the upper bound can be attributed to several factors, including differences in viewing angle, loss of angular momentum as low- and intermediate-mass stars evolve down their convective tracks, and depending on the mechanism that determines the initial angular momentum, possibly to differences in stellar properties (e.g., magnetic field strength, accretion rate during the stellar assembly phase).

Note that we have included in this plot young stars in the mass range 3-8 {\msun} for which {\vsini} values are reported by WSH. Such objects were excluded from Figure 1 on the grounds that these objects are considerably removed from their birthline locations. We include them here because it {\it appears} that stars in this mass range likely conserve angular momentum as they evolve.  We do not expect intermediate mass stars to have magnetic fields or strong winds to carry away angular momentum (see also WSH). If this reasoning is correct, then the angular momenta reported here {\it may} accurately reflect their initial values.

\section{Summary}

Data for O-type stars have been combined with data in the literature to show that over a mass range of a factor of 250 (0.2-50 {\msun}), the specific angular momentum J/M of stars varies slowly and continuously with mass ($J/M \sim M^{0.3}$).  Nearly all stars in this mass range are rotating at rates that are no more than 30\% of the critical velocity calculated from models of stars along the birthline. We conclude that a single mechanism must be at work to keep rotation rates low and at similar values for stars of all masses at birth, this despite the rapid accretion of high angular momentum material from a disk during the stellar assembly phase.   

In the context of MMA and CM models, our results would appear to rule out CM models, which we expect naively to produce more rapid rotation were massive stars formed through mergers. If such mergers take place, they would appear to be the exception rather than the rule. Rather, the continuity of angular momentum properties across the whole mass range from M stars to O stars argues for a common formation mechanism to masses as high as 50{\msun} and adds one more piece to the mounting evidence  (cf WSH) that magnetically-mediated  accretion through a disk is the main mechanism by which stars of all masses form. 

With the present data, we cannot yet extend this conclusion to still higher masses.  We note, however, that Penny (1996) observed several stars with estimated masses larger than 60 {\msun} and found none with {\vsini} $>$ 200 km/sec (well below the estimated escape velocity of $>$ 800 km/sec). It is unclear whether this means that no massive stars rotate extremely rapidly or whether their strong stellar winds have carried away significant amounts of angular momentum.  Observations of very young stars with well established ages are needed to determine the angular momentum properties of stars in the range 50-100 {\msun}.

\acknowledgements 
We thank Diane Harmer of NOAO, who provided generous assistance both in preparing for our WIYN-Hydra observing run and at the telescope, and the referee, Georges Meynet, for his thorough reading of the manuscript and a number of helpful suggestions.

\clearpage
\begin{figure*}[tbp]
\epsscale{0.5}
\plotone{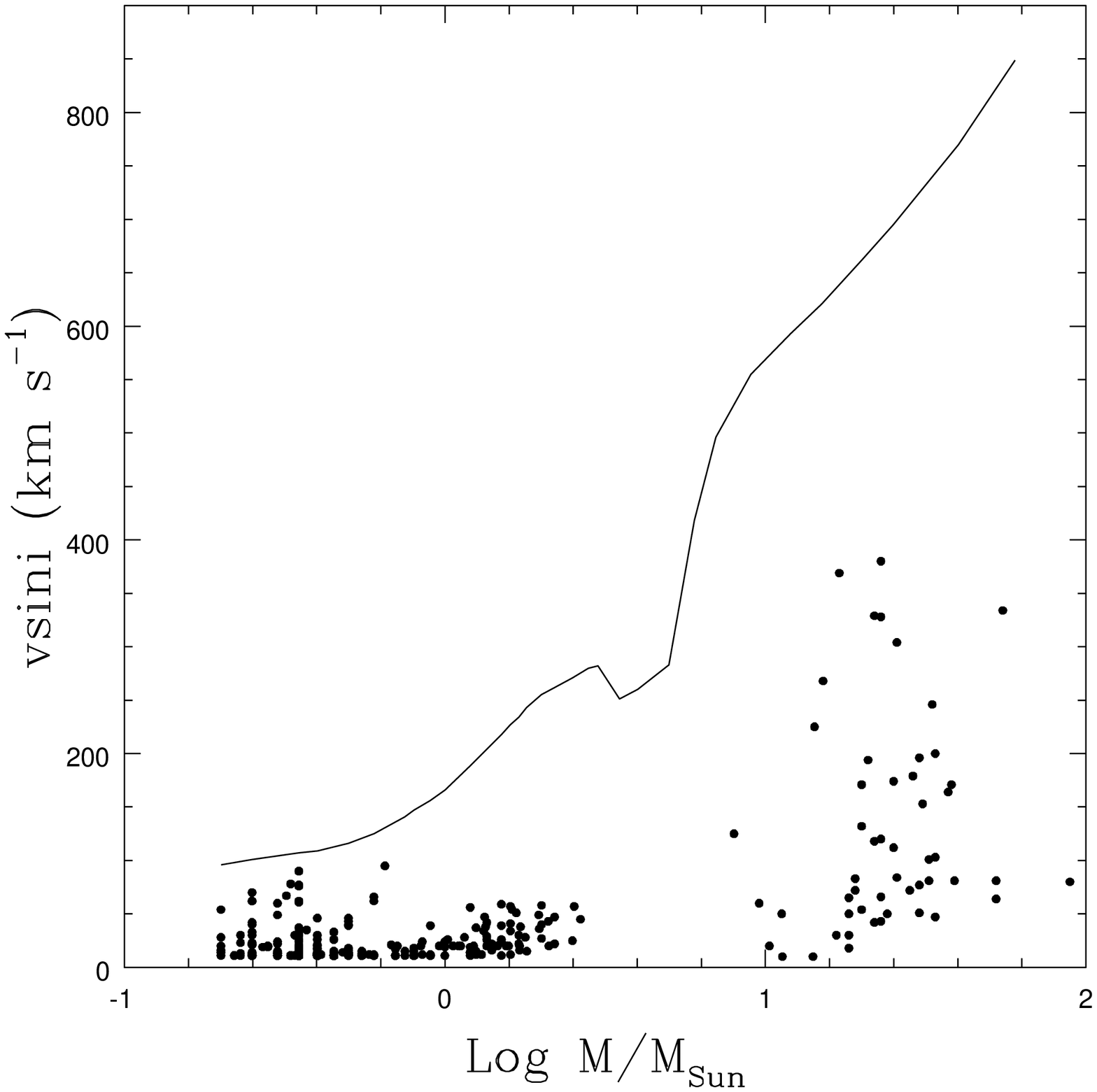}
\caption{Plot of {\vsini} as a function of mass.  The solid line 
indicates the critical velocity $v_{c}$ at the time that 
accretion stops if stars are formed with an 
accretion rate of $10^{-5}$ {\msun} yr$^{-1}$.  For this accretion
rate, stars with M $>$ 8 {\msun} are on the ZAMS when accretion stops, and
$v_{c}$ was calculated from ZAMS models.  For stars with $M<7$ {\msun}, 
$v_{c}$ was calculated from PMS models (Swenson et al. 1994; Siess \&
Forestini 2000).  The decrease in $v_{c}$ at 3 {\msun} is the result of
the rapid expansion of PMS stars associated with deuterium shell burning
(Palla \& Stahler 1992).  Note that most stars lie well below 30\% of
breakup; the median value of {\vsini} lies near 0.1 $v_{c}$ (see also Table 2).}
\label{fig:vesc}
\end{figure*}
\clearpage
\begin{figure*}[tbp]
\epsscale{0.5}
\plotone{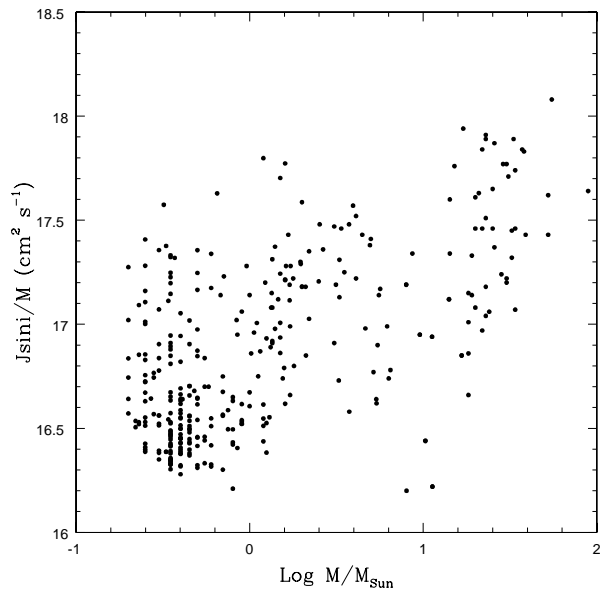}
\caption{Plot of the specific angular momentum {Jsini/M} as a function of stellar
mass.  Note that the upper bound of {Jsini/M} increases slowly with mass with 
no evidence of a discontinuity between the low and high mass stars.  Such
a discontinuity might be expected if the formation mechanism for high 
mass stars were fundamentally different (i.e., via mergers, or CM as defined here) from the formation
mechanism for low mass stars (magnetically-mediated accretion, or MMA).}
\label{fig:jsini}
\end{figure*}
\clearpage

\begin{deluxetable}{rrclrrcccccc}

\rotate

\tablewidth{0pt}
\setlength{\tabcolsep}{0.03in}
\tablecaption{Stellar Data}

\tablenum{1}

\tablehead{\colhead{Cluster} & \colhead{Star Name} & \colhead{WEBDA} & \colhead{RA(2000)}& \colhead{DEC(2000)} & \colhead{Sp.Type} & \colhead{log T$_{eff}$} & \colhead{log L$_{bol}$} & \colhead{Mass} & \colhead{\vsini} & \colhead{log J/M} \\ 
\colhead{} & \colhead{} & \colhead{} & \colhead{} & \colhead{} & \colhead{} &  \colhead{(K)} & \colhead{(L$_\sun$)} & \colhead{(M$_\sun$)} & \colhead{(km s$^{-1}$)} & \colhead{(cm$^2$ s$^{-1}$)} }

\startdata
IC 1805 & -- & 21 & 2 29 30.47 & +61 29 44.2 & O9.5Vf & 4.49 & 4.94 & 22 & 329 & 17.84 \\
IC 1805 & BD+60 499 & 118 & 2 32 16.75 & +61 33 15.0 & 09.5Vf & 4.49 & 4.74 & 18 & 50 & 17.01 &  \\
IC 1805 & BD+60 501 & 138 & 2 32 36.28 & +61 28 25.6 & O7Vf & 4.56 & 5.15 & 29 & 179 & 17.77 &  \\
IC 1805 & BD+60 513 & 232 & 2 34  2.53 & +61 23 10.8 & O8Vf & 4.53 & 5.17 & 26 & 304 & 17.87 &  \\
NGC 2244 & HD46056 & 84 & 6 31 20.86 & +04 50 03.8 & O8Vf & 4.53 & 5.02 & 23 & 328 & 17.91 &  \\
NGC 2244 & HD46149 & 114 & 6 31 52.53 & +05 01 59.2 & O8.5Vf & 4.52 & 5.22 & 28 & 72 & 17.24 &  \\
NGC 2244 & HD46150 & 122 & 6 31 55.52 & +04 56 34.3 & O6e & 4.58 & 5.75 & 52 & 64 & 17.43 &  \\
NGC 2244 & HD46202 & 180 & 6 32 10.47 & +04 57 59.8 & O9Vf & 4.50 & 4.94 & 22 & 42 & 16.97 &  \\
NGC 2244 & -- & 376 & 6 30 33.33 & +04 41 27.9 & O9Vf & 4.50 & 4.83 & 20 & 54 & 17.08 &  \\
NGC 6611 & -- & 161 & 18 18 30.95 & -13 43 08.3 & O8.5V & 4.52 & 4.92 & 22 & 118 & 17.46 &  \\
NGC 6611 & -- & 166 & 18 18 32.22 & -13 48 48.0 & O8.5V & 4.52 & 4.70 & 19 & 83 & 17.33 &  \\
NGC 6611 & -- & 175 & 18 18 32.73 & -13 45 11.9 & O5.5Vf & 4.60 & 5.38 & 37 & 164 & 17.84 &  \\
NGC 6611 & -- & 188 & 18 18 33.73 & -13 40 58.4 & B0V & 4.48 & 4.43 & 15 & 268 & 17.76 &  \\
NGC 6611 & -- & 205 & 18 18 36.42 & -13 48 02.4 & O5Vf & 4.61 & 5.68 & 52 & 81 & 17.62 &  \\
NGC 6611 & -- & 280 & 18 18 42.80 & -13 46 50.8 & O9.5V & 4.49 & 4.54 & 17 & 369 & 17.94 &  \\
NGC 6611 & -- & 314 & 18 18 45.86 & -13 46 30.9 & B0V & 4.48 & 4.80 & 19 & 72 & 17.14 &  \\
NGC 6611 & -- & 367 & 18 18 52.67 & -13 49 42.6 & O9.5V & 4.49 & 4.59 & 18 & 30 & 16.86 &  \\
NGC 6611 & -- & 401 & 18 18 56.19 & -13 48 31.0 & O8.5V & 4.52 & 5.07 & 24 & 50 & 17.06 &  \\
NGC 6611 & -- & 584 & 18 18 23.63 & -13 36 27.9 & O9V & 4.50 & 4.60 & 18 & 18 & 16.66 &  \\
NGC 6611 & HD168504 & -- & 18 20 34.10 & -13 57 15.7 & O7.5III & 4.54 & 5.10 & 26 & 84 & 17.37 &  \\
Cyg OB2 & 59 & -- & 20 29 23.62 & +41 21 41.7 & O8.5V & 4.52 & 5.35 & 34 & 200 & 17.74 &  \\
Cyg OB2 & 70 & -- & 20 29 31.32 & +41 11 08.8 & O9V & 4.50 & 5.37 & 34 & 47 & 17.07 &  \\
Cyg OB2 & 145 & -- & 20 30 02.56 & +41 18 14.1 & O9.5V & 4.49 & 4.67 & 18 & 65 & 17.15 &  \\
Cyg OB2 & 217 & -- & 20 30 26.58 & +41 16 59.1 & O7IIIf & 4.56 & 5.50 & 39 & 81 & 17.43 &  \\
Cyg OB2 & 227 & -- & 20 30 29.31 & +41 15 22.5 & O9V & 4.50 & 4.91 & 21 & 194 & 17.63 &  \\
Cyg OB2 & 258 & -- & 20 30 40.42 & +41 16 07.3 & O8V & 4.53 & 5.08 & 25 & 174 & 17.65 &  \\
Cyg OB2 & 299 & -- & 20 30 51.26 & +41 14 58.5 & O7.5V & 4.54 & 5.21 & 30 & 196 & 17.77 &  \\
Cyg OB2 & 339 & -- & 20 31 02.63 & +41 13 28.8 & O8.5V & 4.52 & 5.03 & 23 & 66 & 17.18 &  \\
Cyg OB2 & 376 & -- & 20 31 11.75 & +41 14 09.3 & O8V & 4.53 & 4.94 & 23 & 120 & 17.51 &  \\
Cyg OB2 & 390 & -- & 20 31 15.36 & +41 07 26.0 & O8V & 4.53 & 5.31 & 32 & 101 & 17.45 &  \\
Cyg OB2 & 448 & -- & 20 31 25.51 & +41 03 10.1 & O6Vf & 4.58 & 5.41 & 38 & 171 & 17.83 &  \\
Cyg OB2 & 480 & -- & 20 31 29.81 & +41 06 50.8 & O7.5V & 4.54 & 5.29 & 33 & 246 & 17.89 &  \\
Cyg OB2 & 485 & -- & 20 31 30.50 & +41 11 18.6 & O8V & 4.53 & 5.08 & 25 & 112 & 17.46 &  \\
Cyg OB2 & 507 & -- & 20 31 33.35 & +41 07 21.5 & O8.5V & 4.52 & 4.82 & 20 & 171 & 17.61 &  \\
Cyg OB2 & 516 & -- & 20 31 35.55 & +40 58 53.2 & O5.5Vf & 4.60 & 6.22 & 90 & 80 & 17.64 &  \\
Cyg OB2 & 531 & -- & 20 31 38.38 & +41 23 09.1 & O8.5V & 4.52 & 5.31 & 32 & 81 & 17.32 &  \\
Cyg OB2 & 534 & -- & 20 31 38.88 & +41 00 39.7 & O7.5V & 4.54 & 5.18 & 30 & 51 & 17.20 &  \\
\tablebreak
Cyg OB2 & 555 & -- & 20 31 43.20 & +41 25 39.9 & O8V & 4.53 & 5.38 & 34 & 103 & 17.46 &  \\
Cyg OB2 & 601 & -- & 20 31 51.40 & +41 09 06.2 & O9.5III & 4.49 & 5.30 & 30 & 77 & 17.22 &  \\
Cyg OB2 & 696 & -- & 20 32 11.66 & +41 07 14.2 & O9.5V & 4.49 & 5.02 & 23 & 380 & 17.89 &  \\
Cyg OB2 & 716 & -- & 20 32 16.60 & +40 54 49.7 & O9V & 4.50 & 4.84 & 20 & 132 & 17.46 &  \\
Cyg OB2 & 745 & -- & 20 32 26.13 & +41 24 40.8 & O7V & 4.56 & 5.22 & 31 & 153 & 17.71 &  \\
Cyg OB2 & 771 & -- & 20 32 41.97 & +41 21 21.7 & O7V & 4.56 & 5.84 & 55 & 334 & 18.08 &  \\
NGC 7380 & -- & 31 & 22 47 50.61 & +58 05 12.2 & O8Vf & 4.53 & 4.99 & 23 & 43 & 17.04 &  \\
\enddata




\end{deluxetable}
\clearpage
\begin{deluxetable}{ccc}




\tablecaption{Median Observed Rotational Velocity Compared with the 
Critical Velocity on the Birthline}

\tablenum{2}

\tablehead{\colhead{No. of Stars} & \colhead{M} & \colhead{$v_{obs}/v_{c}$} \\ 
\colhead{} & \colhead{(M$_\sun$)} & \colhead{Median} } 

\startdata
95 & 0.2 to 0.4 & 0.14 \\
89 & 0.4 to 0.8 & 0.10 \\
76 & 0.8 to 2.5 & 0.11 \\
20 & 8 to 25 & 0.13 \\
34 & 25 to 90 & 0.20 \\
\enddata




\end{deluxetable}

\end{document}